\documentclass[12pt,preprint]{aastex}
\long\def\***#1{{\scshape ***#1***}}



\lefthead{SONGAILA}
\righthead{IGM Metal Evolution}

\begin{document}
\title{The Properties of Intergalactic CIV Absorption II:
     Which Systems Are Associated With Galaxy Outflows?}
\author{Antoinette Songaila\altaffilmark{1} 
}
\affil{Institute for Astronomy, University of Hawaii,
  2680 Woodlawn Drive, Honolulu, HI 96822}

\altaffiltext{1}{Visiting Astronomer, W. M. Keck Observatory, which is jointly
  operated by the California Institute of Technology, the University of
  California, and the National Aeronautics and Space Administration}

\email{acowie@ifa.hawaii.edu}


\begin{abstract}
Using the extremely high S/N quasar absorption-line sample described
in the first paper of the series, we investigate which intergalactic
\ion{C}{4} absorption line systems could be directly associated with galactic
outflows at $z=2-3.5$\ from an analysis of the velocity widths of the \ion{C}{4}
absorption line systems. Only about half the systems with a peak
$\tau$(\rm{CIV}) above 0.4 in the $\lambda 1548~{\rm\AA}$ line (roughly a
column density of \ion{C}{4} above about $2\times10^{13}\ $cm$^{-2}$)
have velocity widths large enough to originate in this way, and very few
of the weaker systems do. The median velocity full width
at a tenth max is found to be $50~{\rm km\ s}^{-1}$ for systems with
peak $\tau({\rm CIV})$ in the range 0.1--0.4 and $160~{\rm km\
s}^{-1}$ for systems with a peak $\tau({\rm CIV})$ in the range
0.4--3. We show that this critical value of $\tau({\rm CIV})$ also
separates systems that could be ionized by galaxy-like spectra
from those in which the ionization is clearly
AGN-dominated. Together the results are consistent with a picture in
which almost all the lower column density, and at least half the higher column
density, systems lie in the more general IGM whereas about half of the
higher column density systems could be produced directly by the outflows
and possibly be ionized by their parent galaxies.
\end{abstract}


\section{Introduction}
\label{secintro}

It is well known that \ion{C}{4} enrichment appears to be a ubiquitous
feature of the intergalactic Ly$\alpha$ forest at least down to
$\tau$(Ly$\alpha$) approaching 1 (Cowie et al. 1995; Tytler et
al. 1995; Songaila \& Cowie 1996; Songaila 1998; Ellison et al.\ 2000;
Songaila 2001; Aguirre, Schaye \& Theuns 2002; Schaye et al.\ 2003;
Simcoe, Sargent \& Rauch 2004; Aguirre et al.\ 2004) and possibly even
at lower values (Cowie \& Songaila 1998; Ellison et al.\ 2000;
Aguirre, Schaye \& Theuns 2002), albeit with substantial variation (of
about a dex) in individual \ion{C}{4}/\ion{H}{1} ratios at every
overdensity and with an equally significant trend with overdensity
(Schaye et al.\ 2003). The origin of these metals is still unclear:
some may be being injected into the IGM from galaxies at the redshifts
in question (Steidel, Pettini \& Adelberger 2001; Adelberger et al.\
2003; Adelberger et al.\ 2005) while others may have been put in place
by earlier galaxy formation or by generations of population III stars
(Gnedin \& Ostriker 1997; Cen \& Ostriker 1999; Wasserburg \& Qian
2000; Madau, Ferrara \& Rees 2001; Qian, Sargent \& Wasserburg 2002;
Bromm, Yoshida \& Hernquist 2003; Venkatesan \& Truran 2003; Mackey,
Bromm \& Hernquist 2003; Fujita et al.\ 2004; Daigne et al.\ 2004;
Yoshida, Bromm \& Hernquist 2004).

 This issue of the fraction arising from current injection versus that
  from early star formation has been brought into prominence by two
  sets of observations.  The first is the invariance of the column
  density distribution and the total $\Omega$(CIV) as a function of
  redshift which we have discussed in Paper I (Songaila 2005; see also
  Songaila 1997, 2001; Pettini et al.\ 2003; Schaye et al.\ 2003).
  These observations show directly that a substantial amount of metals
  was in place prior to $z=5$ and also that subsequent injection
  processes and changes in the ionization environment have not greatly
  changed the total amount of \ion{C}{4} or the way it is distributed
  in column density.  The second is the observation by Adelberger and
  collaborators that there is a strong correlation between the
  \ion{C}{4} absorbers and the Lyman break galaxies (LBGs) at
  redshifts near 3 which they argue suggests that some part of the
  \ion{C}{4} absorption may arise, at least in part, from the outflows
  from these galaxies (Steidel, Pettini \& Adelberger 2001; Adelberger
  et al.\ 2003; Adelberger et al.\ 2005).  However, Porciani \& Madau
  (2005) show that this correlation can be present even if the metals
  are injected from early ($z = 6-12$) small galaxies.  In this
  interpretation the correlation arises because both galaxies and the
  intergalactic absorption lines are biased to overdense regions.
  There is also still substantial statistical uncertainty in the
  galaxy-\ion{C}{4} correlation function (Adelberger et al.\ 2005).

 Higher column density systems, including those associated with damped
Ly$\alpha$ systems, have long been believed to lie in the extended
disks or halos of galaxies.  Indeed, Adelberger et al.\ 2005 find that
roughly a third of \ion{C}{4} absorbers with $N({\rm C~IV}) >
10^{14}~{\rm cm}^{-2}$\ are directly associated with LBGs at $z = 2$.
However, the lower column density \ion{C}{4} systems have been
modelled as originating in the more distributed IGM.  Viewed in this
way the results discussed above are less incongruous and the question
may instead be posed as: above what column density does gas associated
with individual galaxies constitute a major part of the \ion{C}{4}
absorption line population?  It is this question we shall attempt to
answer in the present paper by looking at the internal properties of
the \ion{C}{4} absorption line systems, and, in particular, their
velocity and ionization structure.

 We show that the velocity and ionization properties in combination
suggest that there is a transition at a critical column density of
$N({\rm CIV}) \sim 2\times 10^{13}~$cm$^{-2}$.  Below this column
density most of the absorbers are not currently being ejected by the
galaxies but above this the fraction of the absorbers that could be
arising in galactic outflows increases. If, as this suggests, a
substantial fraction of the \ion{C}{4} absorbing material is generated in
earlier star formation, it becomes easier to understand the invariance
of \ion{C}{4} with redshift. However, the maintenance of the shape of
the distribution function in the redshift range $z=1.5-5$ is still
surprising in view of the multiple mechanisms that are probably
responsible for creating and ionizing the
absorption lines.

 We first reprise some simple numerical arguments to motivate our
analysis. Throughout we use a cosmology with $H_0=65~{\rm km\ s}^{-1}\ 
{\rm Mpc}^{-1}$, $\Lambda=0.67$ and $\Omega=0.33$.

In order to occupy sufficient volume to produce the
\ion{C}{4} absorption lines seen in the intergalactic medium
the metals streaming from the galaxies must have a
very high terminal velocity. The number of absorption line systems
with column densities greater than $N$
per unit absorption length $dX$ 
is given by
\begin{equation}
n(>N)= {{c} \over {H_0}}\  \sigma n_0 
\end{equation}
where $\sigma$ is the fixed cross section of the bubble of \ion{C}{4} 
absorbing material around each galaxy,
$n_0$ is the comoving number density of the galaxies (e.g.
Burbidge et al. 1977) and $dX$ is defined as
\begin{equation}
dX = { {(1+z)^2\ dz} \over 
       {\left [ (1+z)^2 (1+\Omega_{\rm m}z) -
      z(z+2)\Omega_{\lambda}\right ]^{0.5} } }\quad .
\end{equation}

For the adopted cosmology the total observed comoving number density
of LBGs above an R magnitude of 27 at $z=3$ is $\approx
5\times10^{-3}$ Mpc$^{-3}$ (Adelberger \& Steidel 2000). Since
$n(>N)$ is 3.1 for $N=10^{13}~{\rm cm}^{-2}$ at these redshifts we
would require a cross sectional radius of about 200~kpc per LBG to
account for the observed absorption systems above this column density.
Then, since the age of the universe is approximately 1 Gyr at this
redshift, we must have a minimum expansion velocity of over 200 km
s$^{-1}$ in the material producing the \ion{C}{4}
absorbers. Clustering, partial covering in the bubble of expanding
material, or shorter lifetimes would all increase the required
velocity.  It is the coincidence of this required terminal velocity
with the observed expansion velocities in the LBGs that has motivated the
idea that the \ion{C}{4} systems could arise in this way.

The above discussion assumes that every line of sight through the
bubble samples \ion{C}{4} absorbing material, which appears plausible if
there is considerable entrained material in the wind. If the bubbles
were patchy the radius would have to be larger to give the same cross
section, which would begin to stretch the required velocity relative
to that seen in the LBGs. However, if the \ion{C}{4} is ubiquitous in the
bubble we will see it over the line-of-sight velocity range of the
projected line-of-sight through the bubble. For a spherical geometry
the average line width will then be 1.3 times the terminal velocity,
and over most of the face of the bubble the line width will be close
to this value.

Therefore it appears that \ion{C}{4} absorption lines arising in the galaxy
outflows should show large velocity widths of about 300-400 km
s$^{-1}$. Galaxies with lower outflow velocities would have too small
a cross section to produce significant \ion{C}{4} absorption along a typical
line of sight. Furthermore, since there are more low column density
\ion{C}{4} systems, the characteristic velocity width would have to rise with
decreasing column density for the number of weaker systems to be
understood.

Since the bubble absorbers lie close to their host galaxies they may
also experience a proximity effect in which the dominant ionization
field becomes that of the galaxy rather than the metagalactic flux.
At low redshift the metagalactic flux may be AGN-dominated, which
would distinguish it from the galaxy spectral shape. The line ratios
of \ion{Si}{4}/\ion{C}{4} versus those of \ion{C}{2}/\ion{C}{4} are
very different for the two cases and in the second part of the paper
we briefly investigate this alternate diagnostic.

\section{Velocity structure}
\label{secvel}

Motivated by the above discussion we first examine the velocity widths
of the \ion{C}{4} absorption systems to determine which systems are wide
enough to be understood as galaxy outflows and to determine if there
is any dependence of the velocity width on the column density of the
\ion{C}{4} absorbers.

For this analysis we use the cleaned $\tau$(CIV)1548, defined in
paper~I, which is the measured $\tau$(CIV)1548 at all positions where
the \ion{C}{4} doublet ratio is approximately consistent with the expected
value ($\tau$(CIV)1548/$\tau$(CIV)1550 between 1 and 4) and zero
otherwise.  We used the core sample of five ultra-high S/N quasars
described in Paper~I: Q0636+680 ($z_{\rm em} = 3.18$, S/N = 220),
HS0741+4741 ($z_{\rm em} = 3.22$, S/N = 210), Q1422+2309 ($z_{\rm em} =
3.62$, S/N = 340), HS1700+6416 ($z_{\rm em} = 2.72$, S/N = 205),
HS1946+7658 ($z_{\rm em} = 3.03$, S/N = 210).

\begin{figure}[h]
  \includegraphics[width=5in,angle=90,scale=0.9]{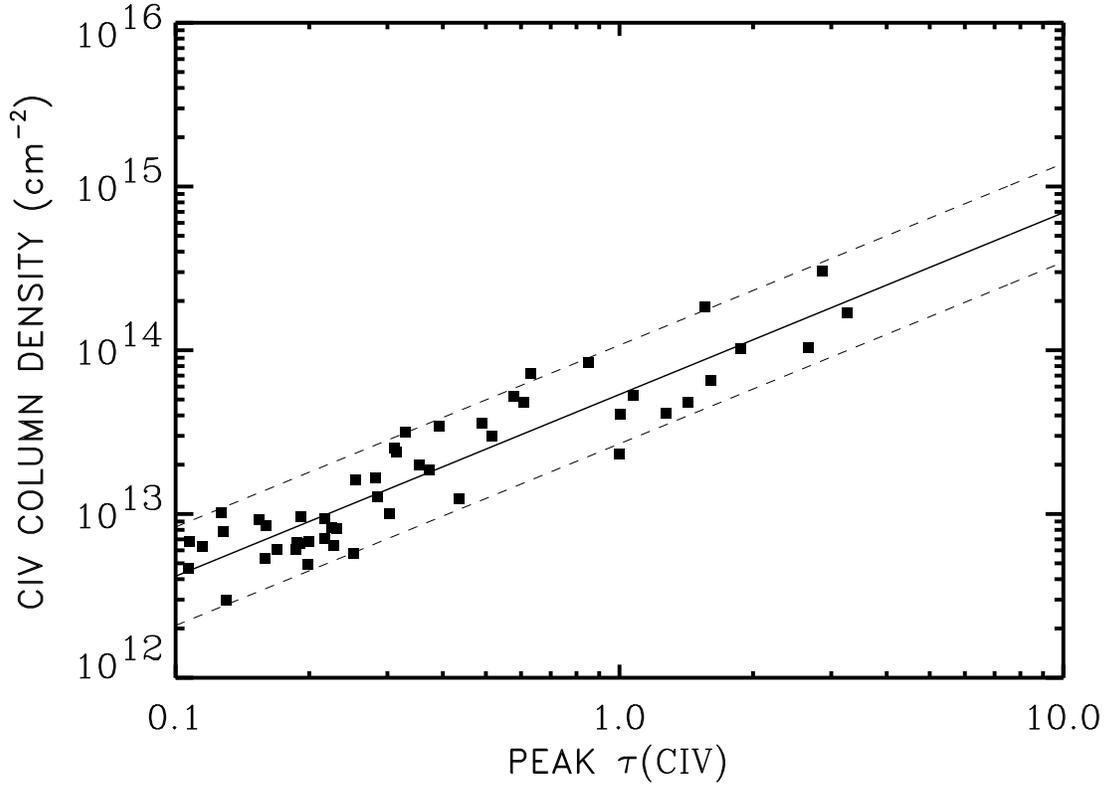}
  \caption{{\it Filled squares\/} : \ion{C}{4} column density determined from
  Voigt profile fitting versus the peak
  optical depth of the \ion{C}{4} $1548$ line in the system. {\it Solid
  line\/} : the power law fit $N({\rm CIV})=5.4\times10^{13}\
  \tau({\rm CIV})^{1.11}\ {\rm cm}^{-2}$.  {\it Dashed lines\/} : a
  multiplicative range of 2 about this fit.
\label{fig:taucol}
}
\end{figure}

In making this analysis we need to characterise the absorption system
with a parameter that is not dependent on the velocity structure of
the system. In particular, high column density systems may be
preferentially selected to be wide in velocity space. We have chosen,
therefore, to measure the strength of the system by using the largest
optical depth in the \ion{C}{4} 1548 line, which we refer to as the
peak optical depth, $\tau_p$.  In fact $\tau_p$ is tightly related to
$N({\rm CIV})$ determined from Voigt profile fitting, as we show in
Figure~\ref{fig:taucol}. The ratio $N({\rm CIV})/\tau_p$\ is
proportional to the velocity integral of $\tau$ divided by the peak
optical depth and so measures the optical depth-weighted velocity
spread in the system (i.e. the total velocity width over which the
bulk of the material arises).  Empirically, the column density is
related to the peak $\tau_p$ as
\begin{equation}
N({\rm CIV})=5.4\times10^{13} \tau_p^{1.11}\ {\rm cm}^{-2}\quad ,
\end{equation}
meaning that the higher column density absorbers are systematically
wider than the lower column density ones, though the rise is not
rapid. Nearly all the absorbers fall within a multiplicative factor of
two of this relation so the dispersion of the optical-depth-weighted
velocity widths at any given $\tau_p$ is not large.

For the present analysis we want a quantity that is a measure of the
full spread in velocity of each system and that is not a function of
the strength of the system. We choose to define the width of the
system as the maximum difference in velocity between positions in the
system that satisfy the condition that $\tau$ is greater than some
fraction of the peak $\tau$. This definition does not require that
there be continuous absorption above this value throughout the
velocity range and therefore we must restrict the search to some
finite velocity range.  We choose to search to positions that lie
within $350~{\rm km\ s} ^{-1}$ of the peak $\tau$, which in turn
restricts the maximum measurable width to a value of 700 km s$^{-1}$.
This is appropriate if we want to look at the velocity spread of
patchy material spread throughout an outflow.  The procedure will
occasionally connect neighboring absorption line systems which are not
in fact causally connected.  This will then result in a large
overestimate of the velocity width for the occasional individual
system.  Such contamination is more probable at lower $\tau$ where
there is a higher density of systems and must be dealt with
statistically by modelling the number of chance connections that
occur.

\begin{figure}[h]
  \includegraphics[width=5.0in,angle=90,scale=1.0]{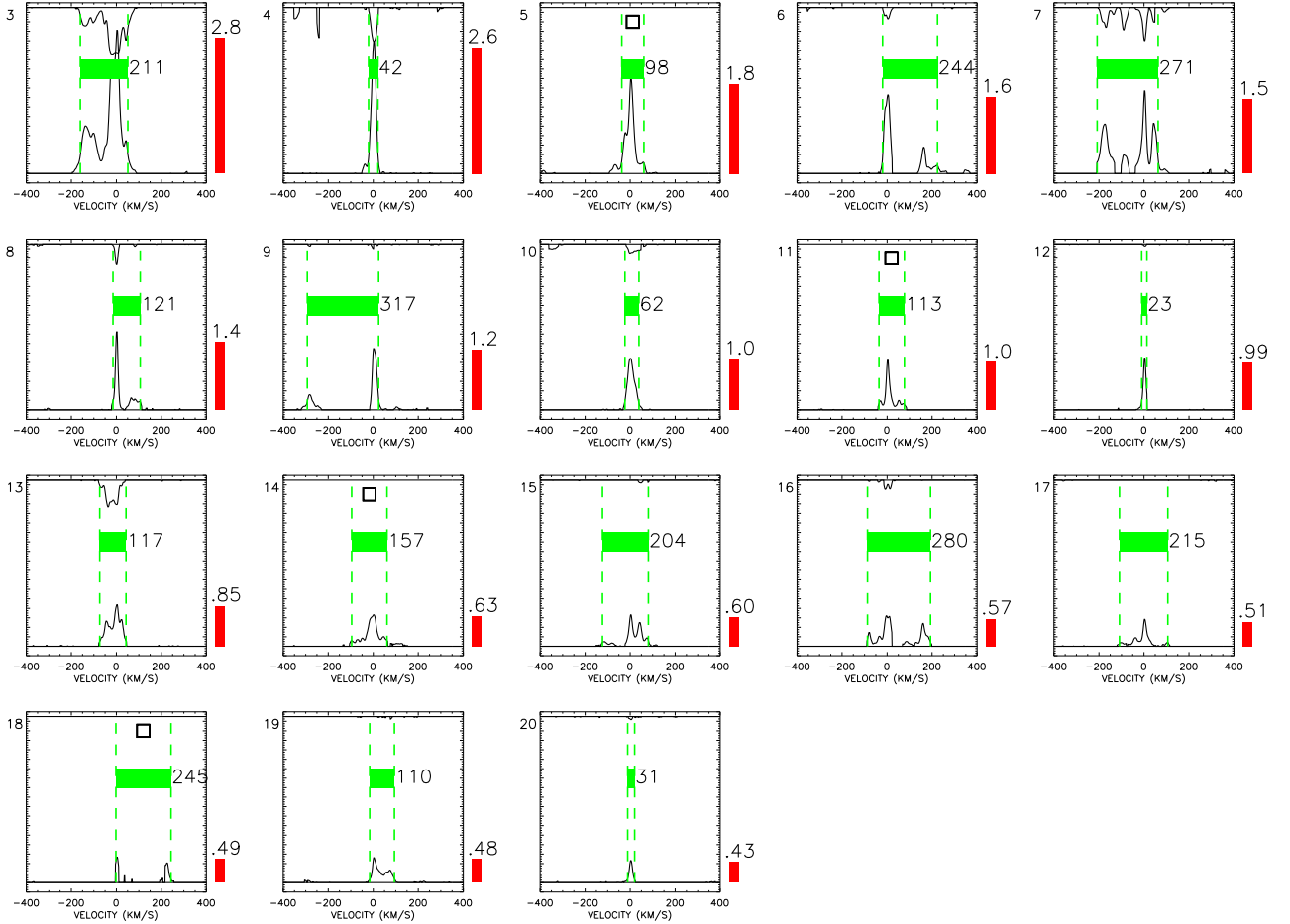}
  \caption{C~IV systems from Table 1, with peak C~IV optical depth
  above 0.4.  Each thumbnail corresponds to
  one system from the Table, indexed by the small numeral in the upper
  left.  The lower
  plot is C~IV optical depth as a function of velocity, in km\
  s$^{-1}$,  for the
  automatically detected system, with zero velocity marking the
  position of peak optical depth.  The upper curve is the corresponding
  Si~IV flux, normalized to unity, with open squares marking systems
   where Si~IV absorption
  is in the Ly~$\alpha$\ forest.  The horizontal green bar marks the
  automatically determined full width at tenth max of the system
  with endpoints denoted by vertical dashed green lines);  this width, in km\
  s$^{-1}$, is marked to the right of the green bar.  The vertical red
  bar denotes the peak C~IV optical depth whose value is given above
  the bar.
\label{fig:tuf1}
}
\end{figure}

\begin{figure}[h]
  \includegraphics[width=5.0in,angle=90,scale=1.0]{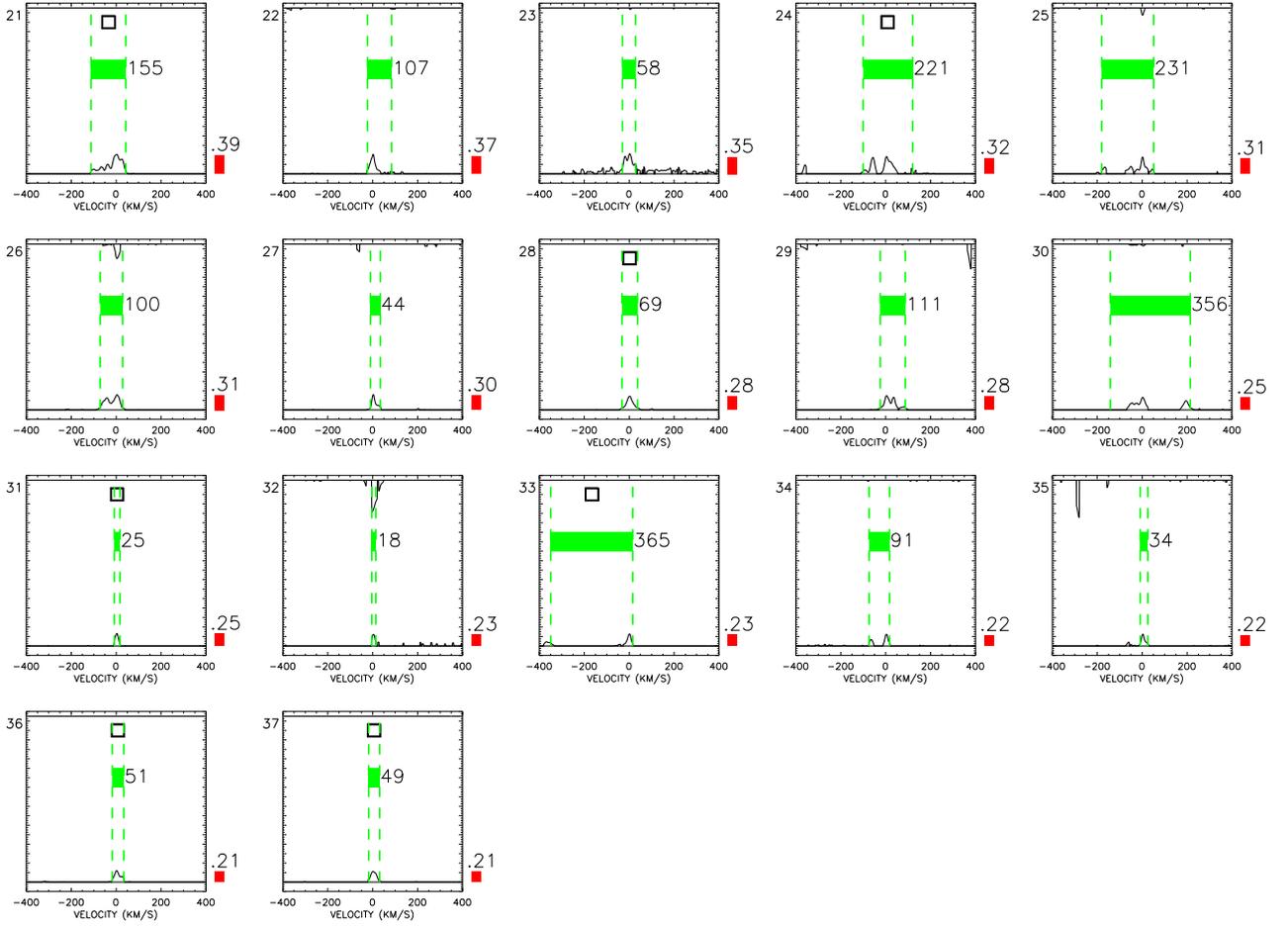}
  \caption{As in Fig.~\ref{fig:tuf1}, for peak optical depth between 0.2 and 0.4.
\label{fig:tuf2}
}
\end{figure}

\begin{figure}[h]
  \includegraphics[width=5.0in,angle=90,scale=1.0]{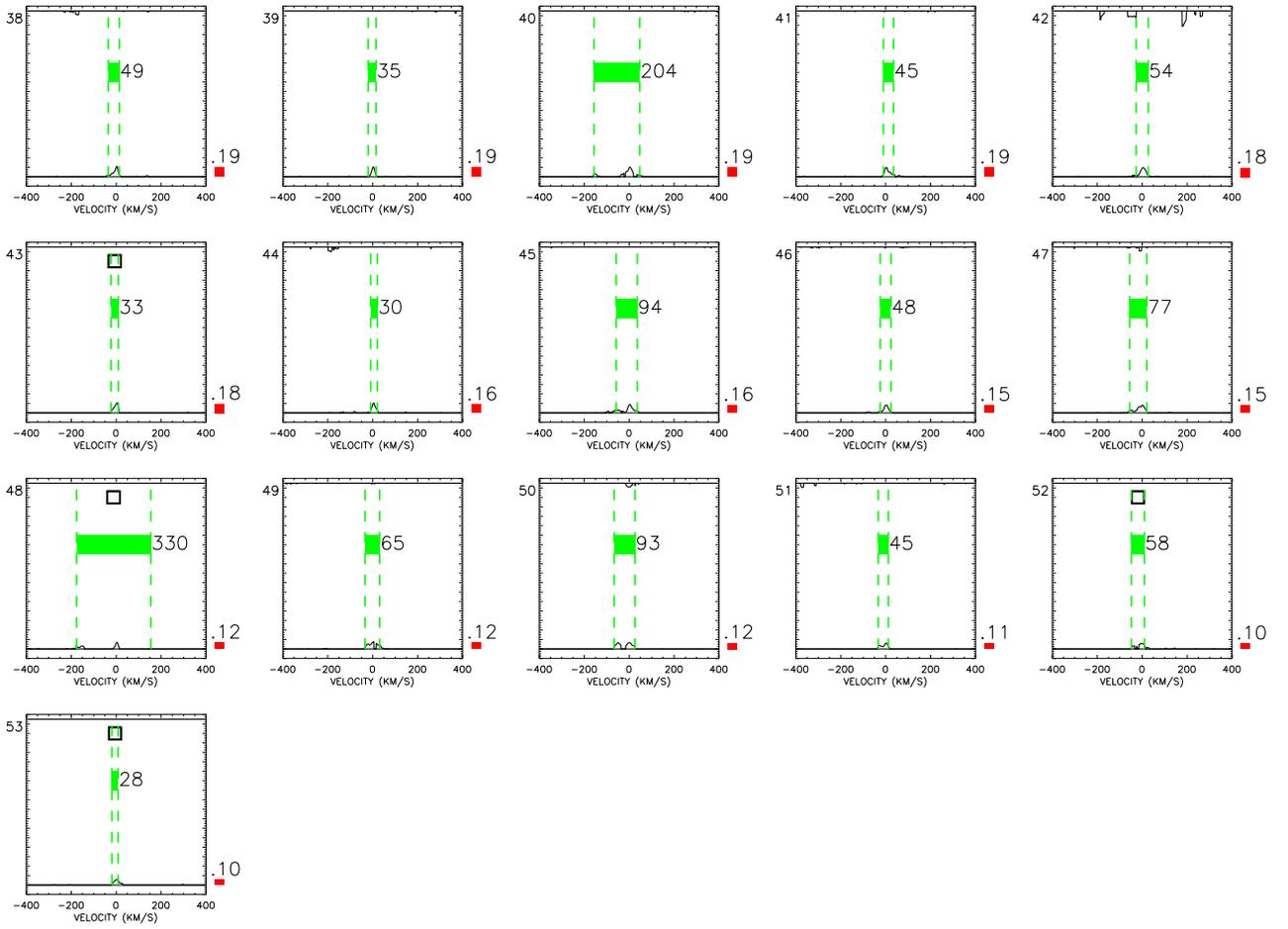}
  \caption{As in Fig.~\ref{fig:tuf1}, for peak optical depth below 0.2.
\label{fig:tuf3}
}
\end{figure}

For each of the 5 quasars we searched a region of the spectrum
longward of the quasar's L$\alpha$, and also more than $10,000~{\rm km\
s}^{-1}$ shortward of the quasar's \ion{C}{4} emission to avoid forest
contamination and \ion{C}{4} systems associated with the quasar itself. We
worked downward in optical depth to select the sytems, restricting
ourselves to $\tau_p$s that were more than two velocity windows from
an existing system. The final sample contains 53 \ion{C}{4} absorption
line systems with a range of $\tau_p$ from 0.1 to 4. The systems range
in redshift from 2.0 to 3.6 with a median of 2.7.  Information on the
53 systems is collected in Table~\ref{tbl:1} which lists index number,
quasar, redshift and value of the automatically-determined peak
optical depth within the system, full width at tenth max (FWTM), and
Voigt profile-fitted values of the system's centroid redshift and
\ion{C}{4}, \ion{Si}{4} and \ion{C}{2} column densities.  The 53
systems are shown in Figures~\ref{fig:tuf1} to \ref{fig:tuf3}.

\begin{figure}[h]
  \includegraphics[width=5in,angle=90,scale=0.9]{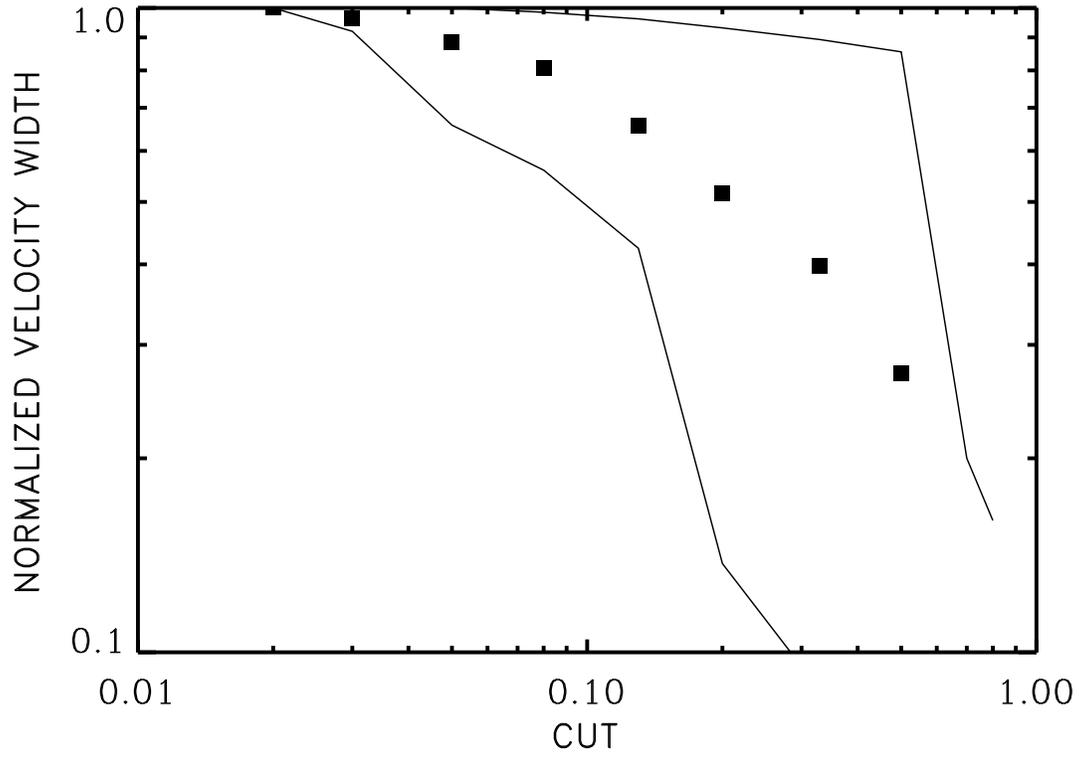}
  \caption{{\it Filled squares\/} : average ratio of the full
  width measured at the specified cut level relative to that measured
  at a cut of 0.02 for the 6 objects with $1.4 < \tau < 3$. 
  {\it Solid lines\/} : maximum and minimum values measured in
  the sample.  \label{fig:cut_up} }
\end{figure}

In order to choose an appropriate cut value for the ratio of the
optical depth to the peak optical depth at which to measure the
velocity width, we analysed the growth in width versus cut value
for the stronger absorption line systems in the sample ($1.4 < \tau_p
< 3$).  This is shown in Figure~\ref{fig:cut_up}.  The measured
velocity widths approach the asymptotic values as the cut level is
reduced, and the average width has reached 85\% of the asymptotic
value at a cut of 0.1. We adopt this value for the analysis, referring
to this as the full width at tenth max (FWTM) of the absorption
system. Cutting at such a low level places stringent demands on the
S/N of the quasar sample and, even for the very high S/N data of the
present quasar sample, restricts us to absorption systems with
$\tau_p$ in excess of 0.1.

\begin{figure}[h]
  \includegraphics[width=2.8in,angle=90,scale=0.9]{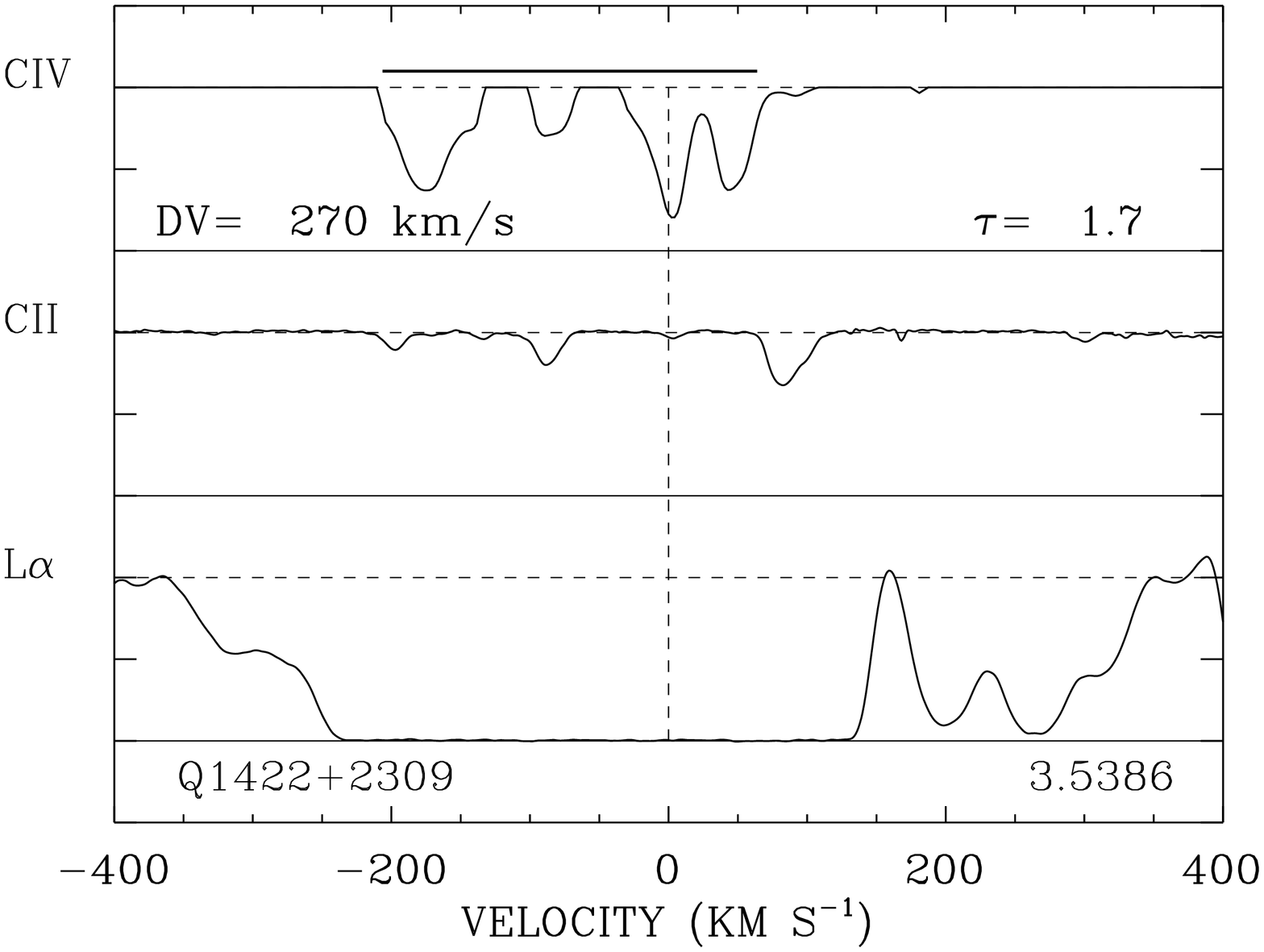}
  \includegraphics[width=2.8in,angle=90,scale=0.9]{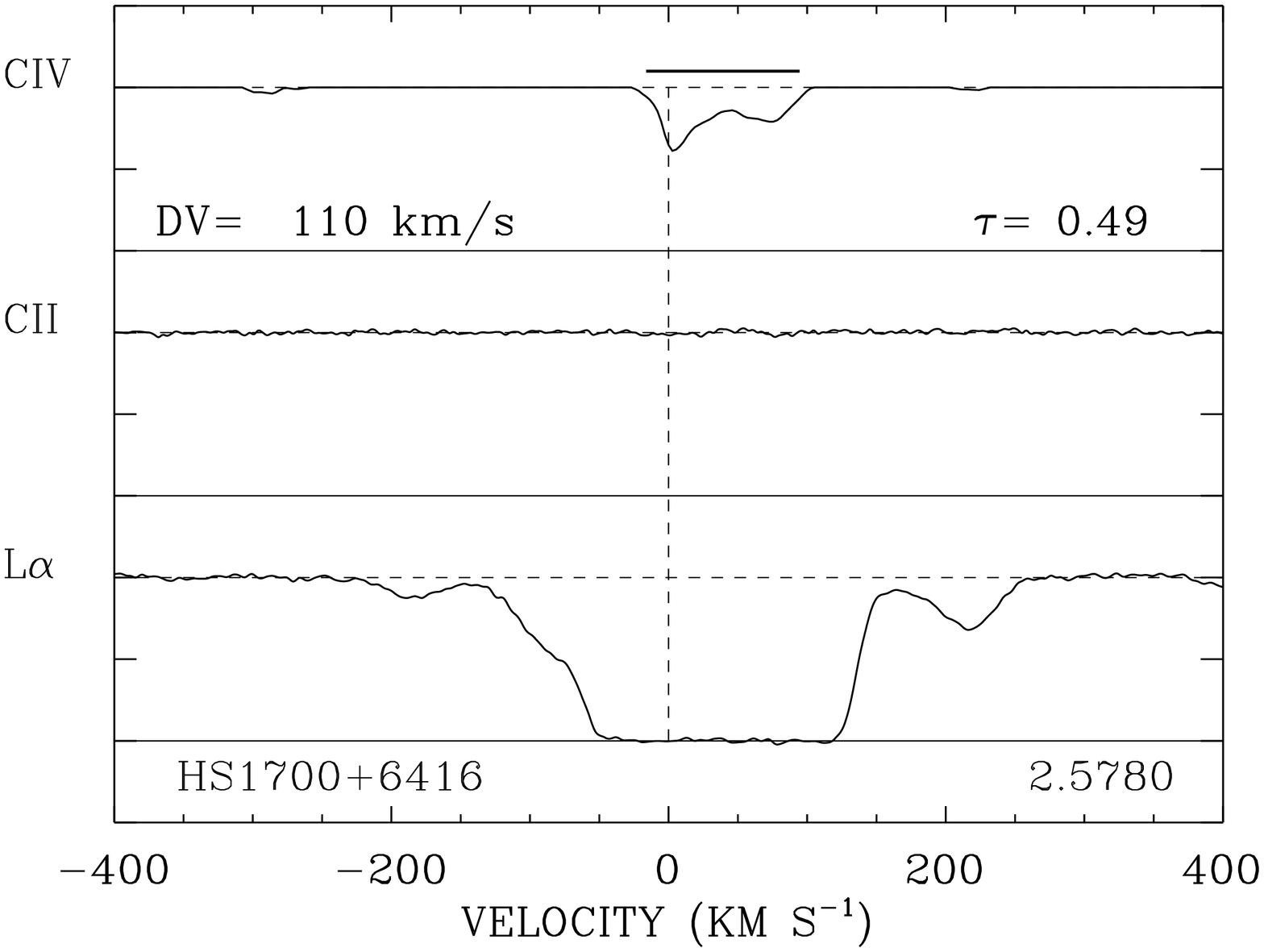}
  \includegraphics[width=2.8in,angle=90,scale=0.9]{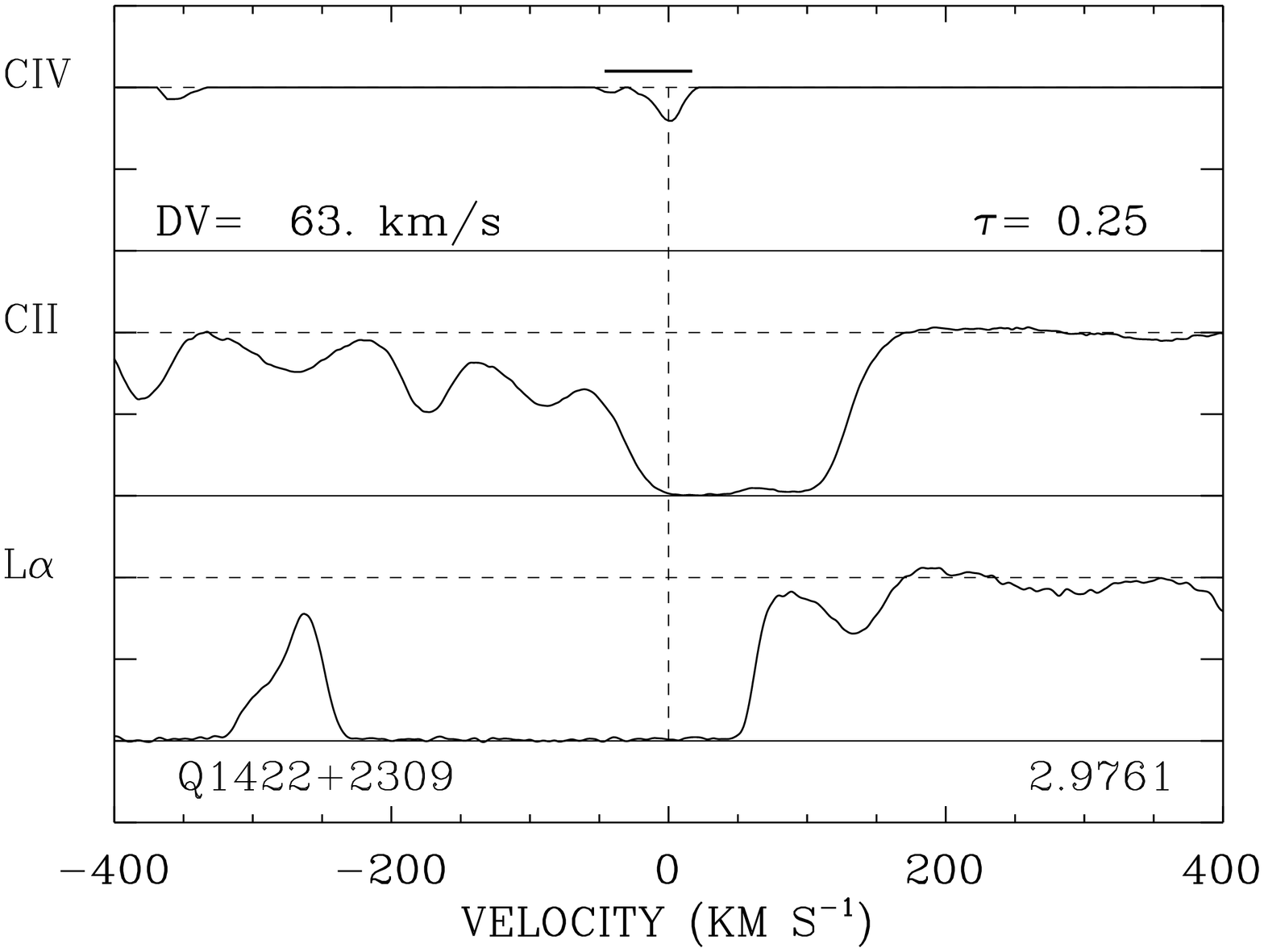}
  \caption{Sample plots showing the width at a tenth of the peak \ion{C}{4}
  optical depth for a range of $\tau_p$. The width is shown at the top
  of each plot over the \ion{C}{4} absorption ({\it solid horizontal
  line\/}). The {\it dashed vertical line\/} shows the position of the peak
  $\tau$. Absorption profiles for CII and L$\alpha$ are also
  shown. Note that for the $z = 2.9761$\ system in Q1422+2309, C~II is in the
  Ly~$\alpha$\ forest.  \label{fig:demo} }
\end{figure}

\begin{figure}[h]
  \includegraphics[width=5in,angle=90,scale=0.9]{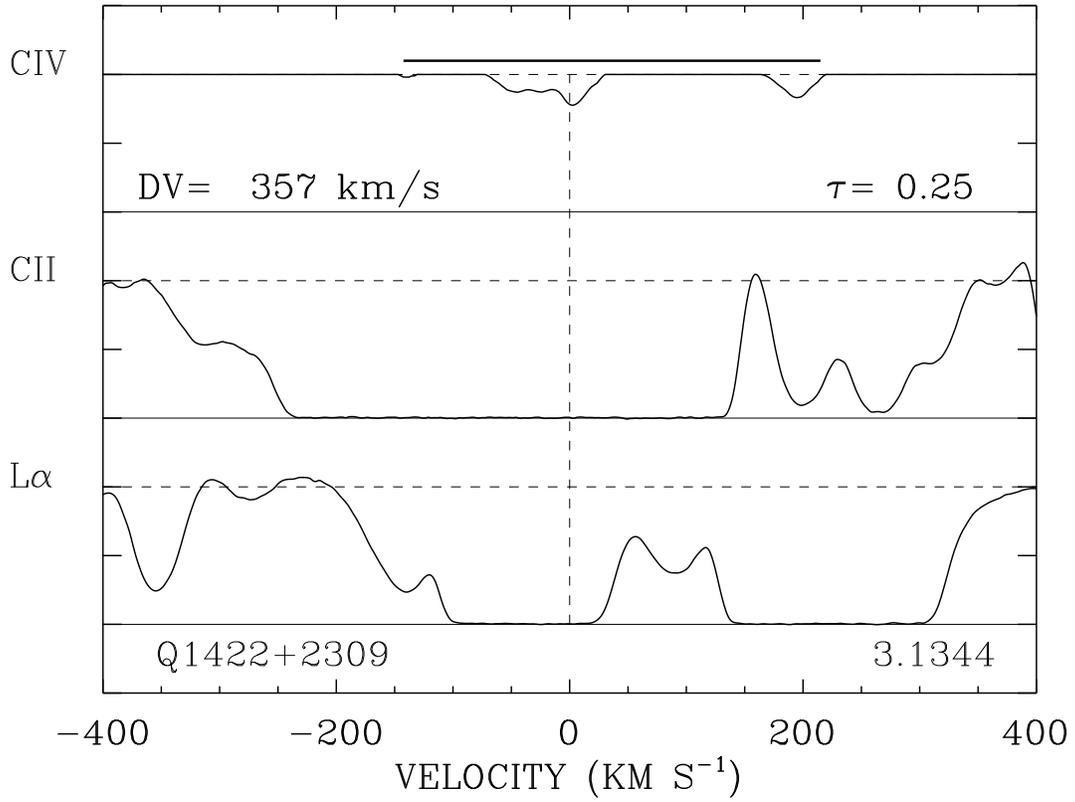}
  \caption{Sample plots showing an example where the
  method may have connected two neighboring systems
  and produced too large a velocity width.  Note that C~II is in the Ly$\alpha$\
  forest for this system.
  \label{fig:false_wid}
}
\end{figure}

Samples of the FWTM measure of absorption line systems with a wide
range in $\tau_p$ are shown in Figure~\ref{fig:demo} where we also show the
corresponding CII and Ly$\alpha$ lines for comparison. In each case
the thick solid line shows the measured width of the system. In nearly
all cases the measured velocity spread lies within the saturated
portions of the corresponding L$\alpha$ line, and where CII absorption
is seen it is generally comparable to or narrower than the velocity
spread seen in the \ion{C}{4} line.
A probable example of a case where the method has
actually joined two independent systems is shown in
Figure~\ref{fig:false_wid}, though in any given case we can only
statistically estimate how likely such a chance projection is.

\begin{figure}[h]
  \includegraphics[width=5in,angle=90,scale=0.9]{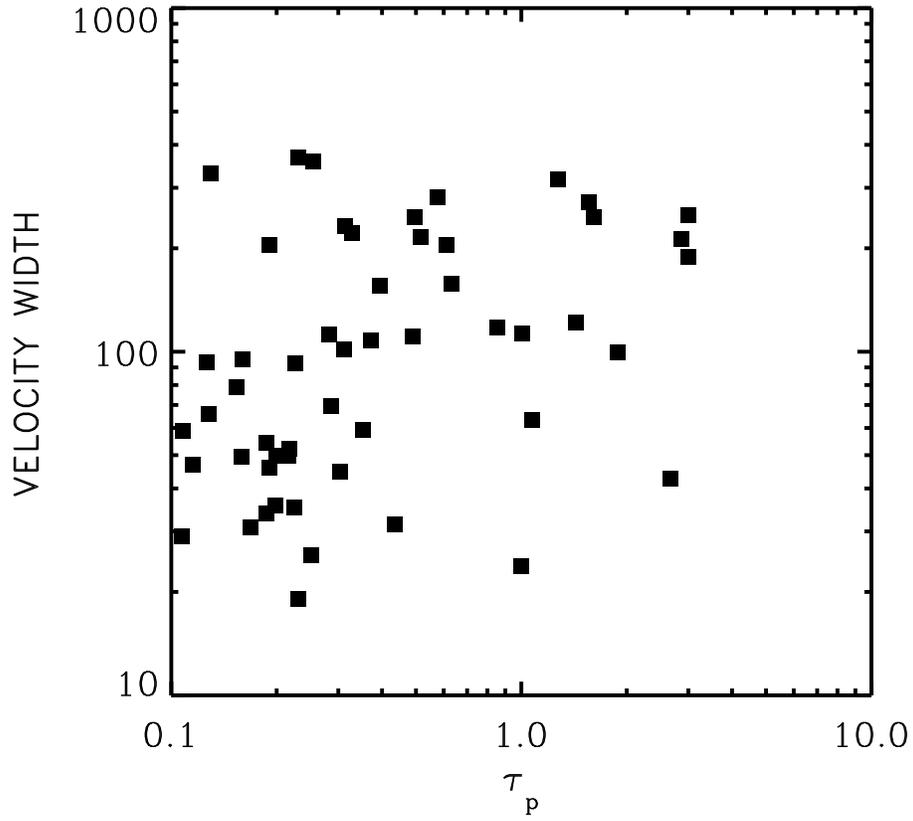}
  \caption{FWTM velocity width determined by the automatic procedure versus 
  peak optical depth in \ion{C}{4} 1548$\rm\AA$. 
  \label{fig:tauvel}
}
\end{figure}

\begin{figure}[h]
  \includegraphics[width=2.8in,angle=90,scale=0.9]{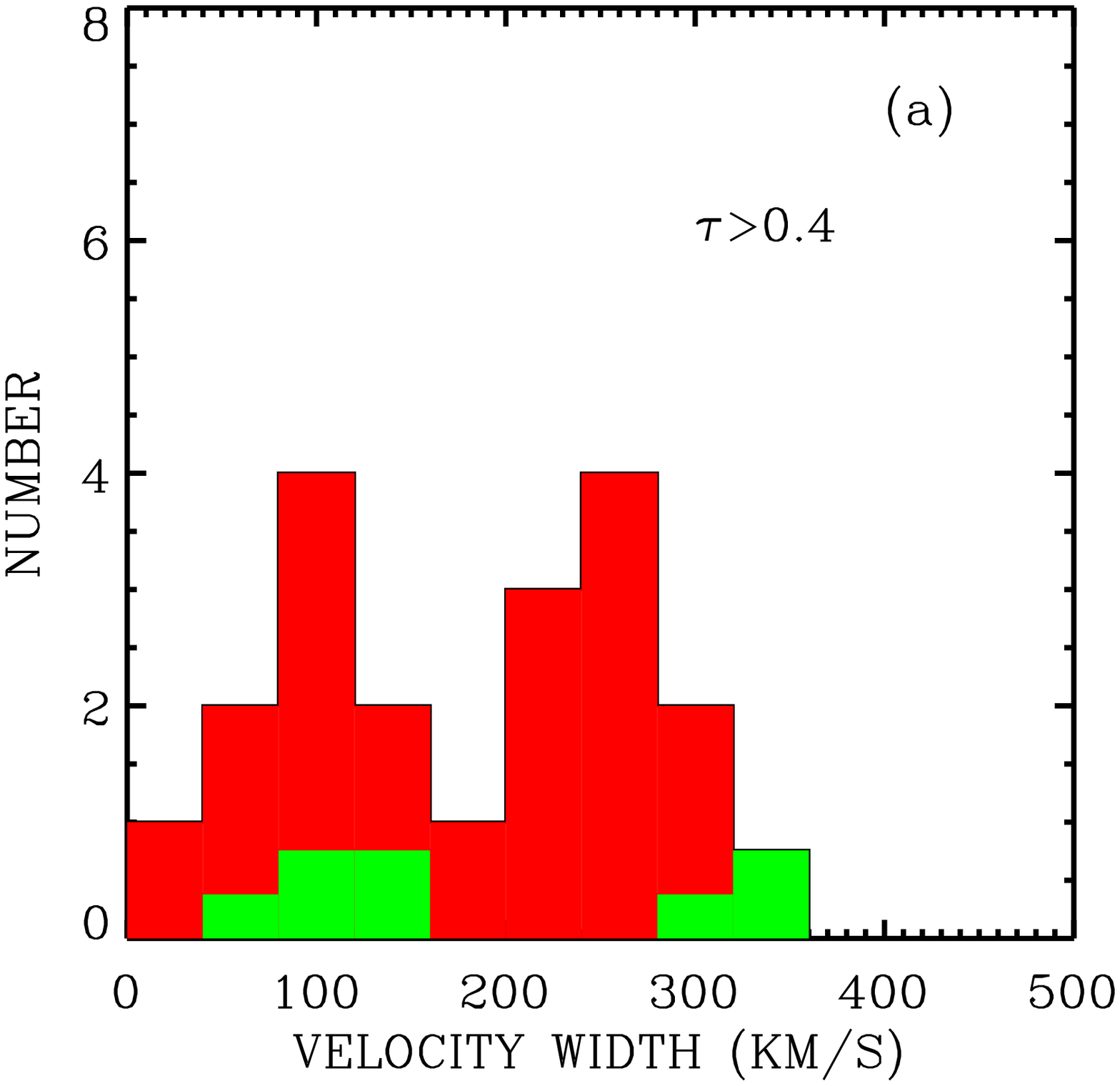}
  \includegraphics[width=2.8in,angle=90,scale=0.9]{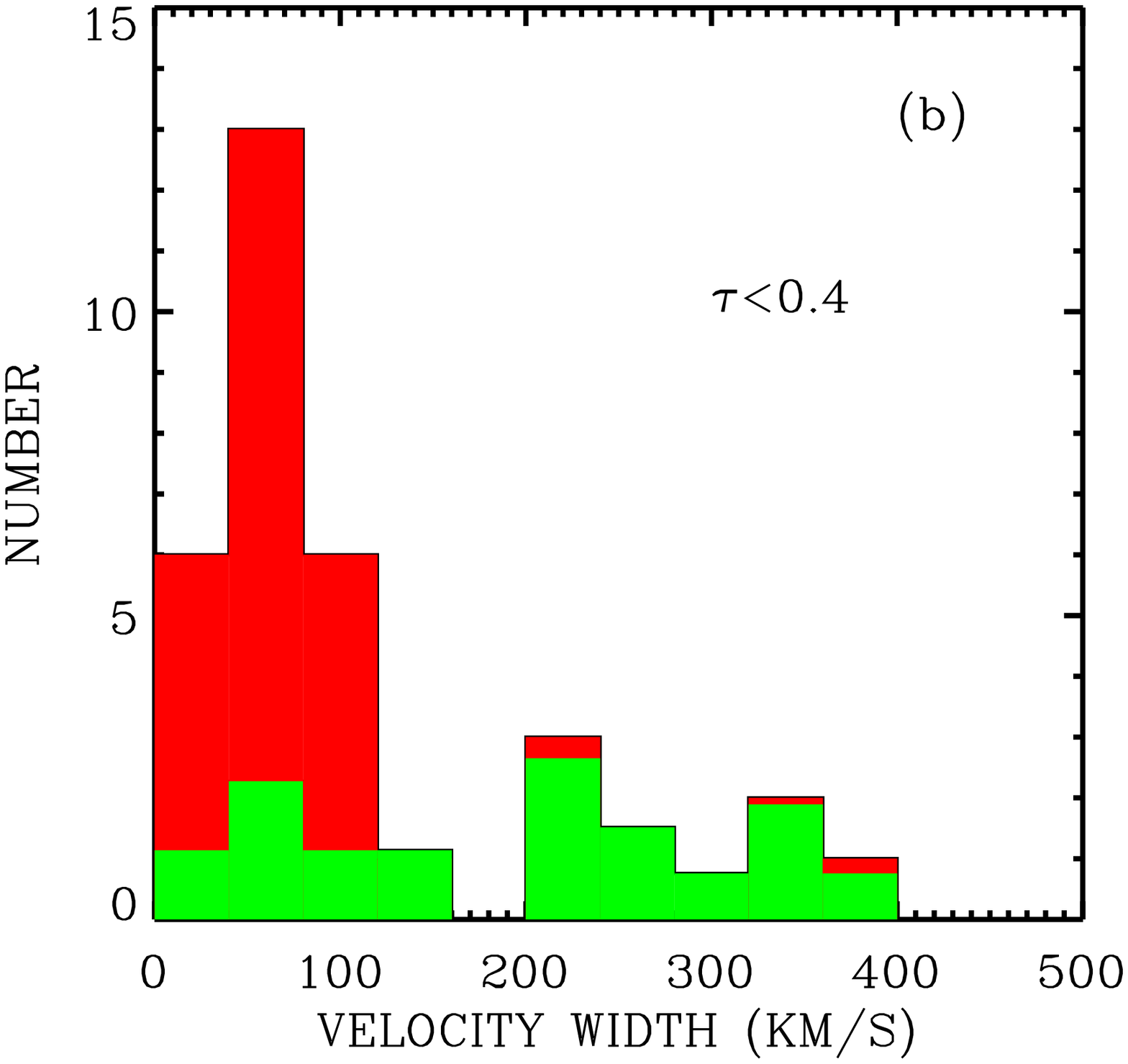}
  \caption{ {\it Panel (a)\/} : histogram of the velocity widths for
  systems with $\tau_p > 0.4$.  {\it Panel (b)\/} : histogram of the
  velocity widths for systems with $\tau_p = 0.1-0.4$. 
  {\it Red (dark) histogram}: measured values;  {\it green (light) histogram}:
  expected number of false systems.
\label{fig:vel-wid} }
\end{figure}

The measured velocity widths for the systems are shown as a function
of $\tau_p$ in Figure~\ref{fig:tauvel} and the corresponding
histograms of the distributions in two $\tau_p$ ranges are shown in
red in Figure~\ref{fig:vel-wid}.  In order to estimate the degree of
contamination we adopted a procedure in which the redshift positions
of the $\tau_p$s were inserted into the spectrum of one of the other
quasars and the FWTM measured around this position using the same
procedure as in the real systems. In most cases
this is zero but contamination results in a number of systems in which
a non-zero width is measured.  The advantage of this procedure over
simulations is that it properly includes the effects of systematics in
the spectra, such as incompletely subtracted sky lines, telluric
absorption, poor continuum fitting, etc. The process produces a false
sample which is slightly less than four times the true sample (since
only overlapping portions of the spectra can be used) and which can be
scaled to estimate the number of contaminating systems in the true
sample. The expected contamination is shown by the green histograms of
Figure~\ref{fig:vel-wid}.

At low $\tau_p$ almost all of the wide systems (${\rm FWTM} > 150~{\rm
km\ s}^{-1}$) are false.  Seven systems are observed, whereas the
analysis of the false sample predicts that there should be 8.7 systems
with random widths of this value.  The $1~\sigma$\ upper limit on the
number of real systems is 2.  At widths less than $150~{\rm km\
s}^{-1}$\ there are 26 systems and we expect 4.9 systems.  Therefore,
fewer than 10\% of the weak \ion{C}{4} absorbers are wide.  The median
width of the distribution corrected for the false cases is 50 km
s$^{-1}$ and the systems are closely distributed around this value. In
contrast, contamination is, as expected, a much less serious problem
at high $\tau_p$ and here there is a much wider spread in the FWTM
velocity width and even a suggestion of bimodality, though the number
of systems is too small for this to be highly significant. The median
value is 160 km s$^{-1}$ and the maximum velocity width is about 300
km s$^{-1}$.

We conclude that at low $\tau_p$ ($< 0.4$) the systems are too narrow
to be plausibly associated with galaxy outflows but that at higher
$\tau_p$ some systems are much wider and it is possible that at least
some fraction of the systems are formed in this way.  The dividing
$\tau_p$ of about 0.4 corresponds to a column density $N({\rm CIV})$
of $2\times 10^{13}$ \ cm$^{-2}$.  We emphasize that the data are of
sufficiently high quality that such systems could be detected easily. 

\section{Ionization balance}
\label{secion}

 A luminous Lyman break galaxy at $z=3$ with an observed $AB$\
magnitude of about 24 at the redshifted Lyman continuum would produce
an outgoing ionizing flux of about $4 \times 10^{-22}\ (\phi/0.01)$\
$(R/200\ {\rm kpc})\ {\rm erg\ cm^{-2}\ s^{-1}\ Hz^{-1}}$\ just below
the ionization edge. Here $\phi$ is the escape fraction of ionizing
photons and $R$ is the radial separation of the absorber from the
galaxy. Within the considerable uncertainty in the escape fraction
(e.g. Steidel, Pettini \& Adleberger 2001; Giallongo et al.\ 2002) the
ionization of associated absorbers could be dominated by the galaxy
or, if the escape fraction is low, by the average metagalactic
flux. Thus it is possible that we might be able to distinguish between
absorbers in galaxy outflows and those in the general IGM based on
their ionization balance.

   The simplest test is to consider the ratio \ion{Si}{4}/\ion{C}{4}
 ion versus \ion{C}{2}/\ion{C}{4}. As is well known (e.g.  Rauch,
 Haehnelt, \& Steinmetz 1997) galactic spectra that cut off at high
 energies produce high \ion{Si}{4}/\ion{C}{4} ratios even in systems
 with high ionization parameters and low \ion{C}{2}/\ion{C}{4}. In
 contrast, with an AGN power law spectrum \ion{Si}{4}/\ion{C}{4} drops
 as \ion{C}{2}/\ion{C}{4} does.

\begin{figure}[h]
  \includegraphics[width=5.0in,angle=90,scale=1.0]{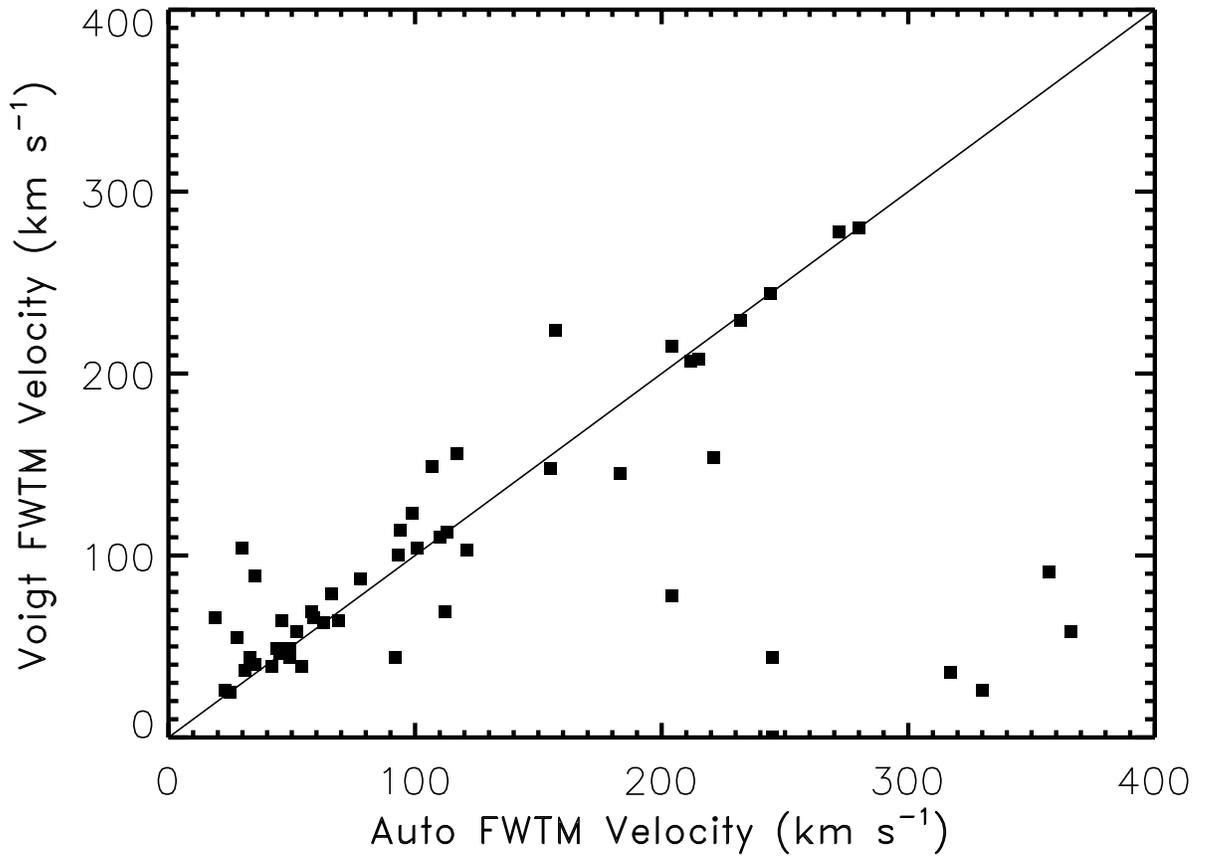}
  \caption{Comparison of the automated measure of FWTM with the same
  quantity measured from Voigt profile fits.  The two measures
  generally agree well except for the small number of cases for which
  the Voigt profile fitting has treated a wide velocity system as two
  separate complexes.
\label{fig:velcomp}
}
\end{figure}

The various ions may have somewhat different velocity structure;  in
particular, \ion{C}{2} is often different from \ion{C}{4}.  For this
reason, we use in this section the Voigt profile-fitted column
densities given in Table~\ref{tbl:1}, which are fitted over the whole
range of visible components, and, for consistency, the FWTM
velocity width measured on the Voigt profile fit.  We show the
comparison of the two FWTM velocities in Figure~\ref{fig:velcomp}.  In
general, the two measures agree extremely well except for the small
number of cases where the Voigt fitting has split a system treated as
one complex by the automatic procedure into two complexes.  

\begin{figure}[h]
  \includegraphics[width=5.0in,angle=90,scale=1.0]{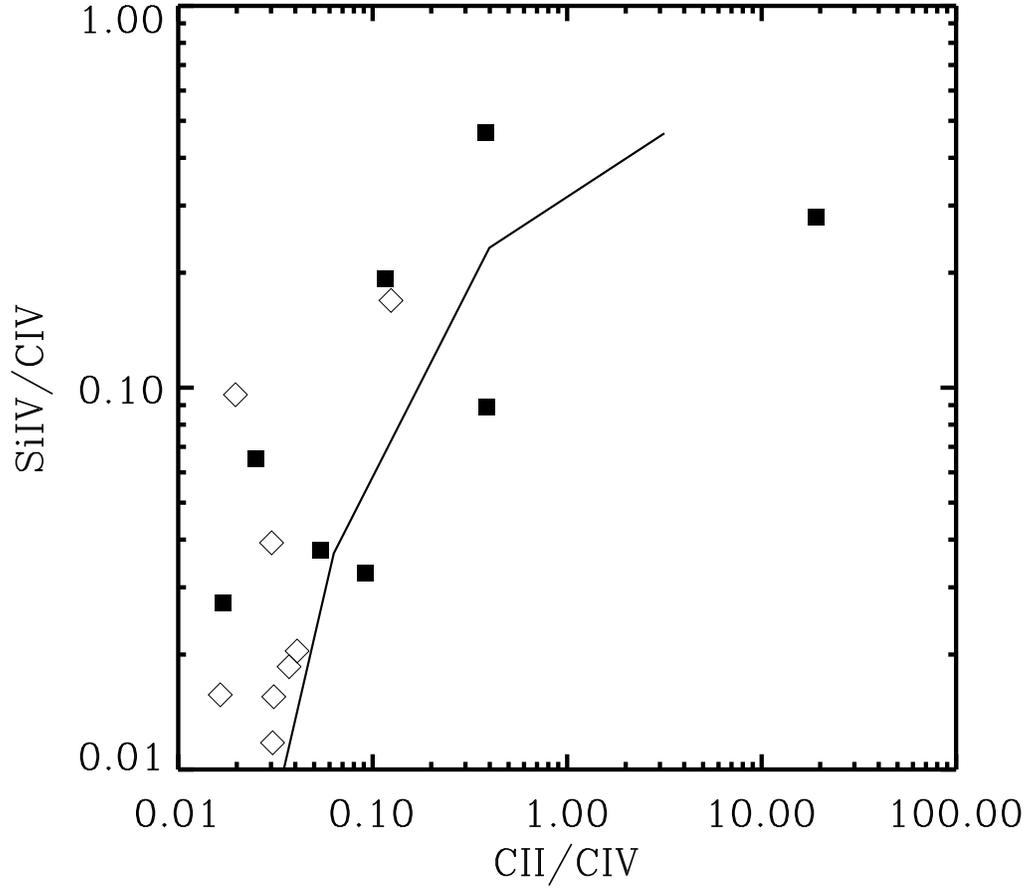}
  \caption{\ion{Si}{4}/\ion{C}{4} versus \ion{C}{2}/\ion{C}{4}
  measured from Voigt profile fitting is compared to a simple
  prediction made using an AGN power law spectrum and computed for a
  Si/C ratio of 2.5 times solar.  {\it Filled squares}:  systems with
  $\Delta v({\rm FWTM}) > 100~{\rm km\ s}^{-1}$.  {\it Open diamonds}:
  systems with $\Delta v({\rm FWTM}) < 100~{\rm km\ s}^{-1}$.
\label{fig:si4c4_c2c4}
}
\end{figure}

In Figure~\ref{fig:si4c4_c2c4} we show the ratios for the 18 systems
 in Table~\ref{tbl:1} where both \ion{C}{2} and \ion{Si}{4} lie above
 the Lyman alpha forest and where the redshift of the aborber is less
 than 3.1 (chosen to roughly lie at the upper boundary of \ion{He}{2}
 reionization). Systems with high velocity spreads based on the Voigt
 fitted FWTM are shown by filled squares while those with $dv({\rm
 FWTM})< 100\ {\rm km\ s^{-1}}$ are shown with open diamonds.  All of
 the systems are roughly consistent with a simple AGN ionization model
 with a Si/C abundance of 2.5 times solar, shown by the solid
 line. Nearly all of the narrow systems have \ion{Si}{4}/\ion{C}{4}
 values that are well below that expected from a galaxy ionizing
 spectrum even with a solar Si/C abundance. We can conclude therefore
 that the narrow systems are ionized by an AGN-dominated metagalactic
 flux. The broader systems, which generally have higher column
 densities, have lower ionization parameters and higher ratios, mostly
 lie in the region where the differentiation is less clear, and,
 particularly in view of possible flexibility in the Si/C abundance,
 could be ionized by either a galaxy spectrum or an AGN one.

\begin{figure}[h]
  \includegraphics[width=2.8in,angle=90,scale=1.0]{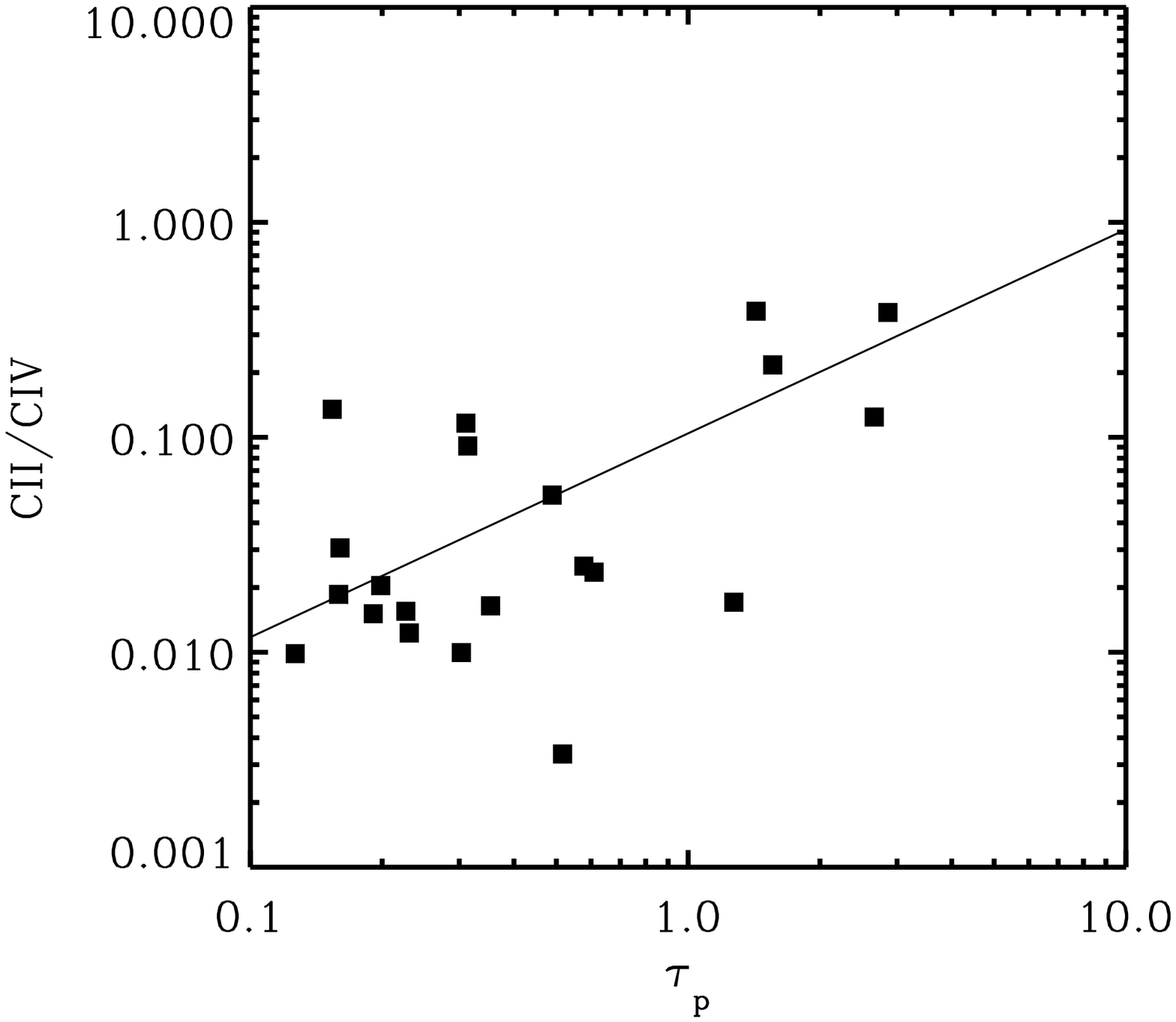}
  \includegraphics[width=2.8in,angle=90,scale=1.0]{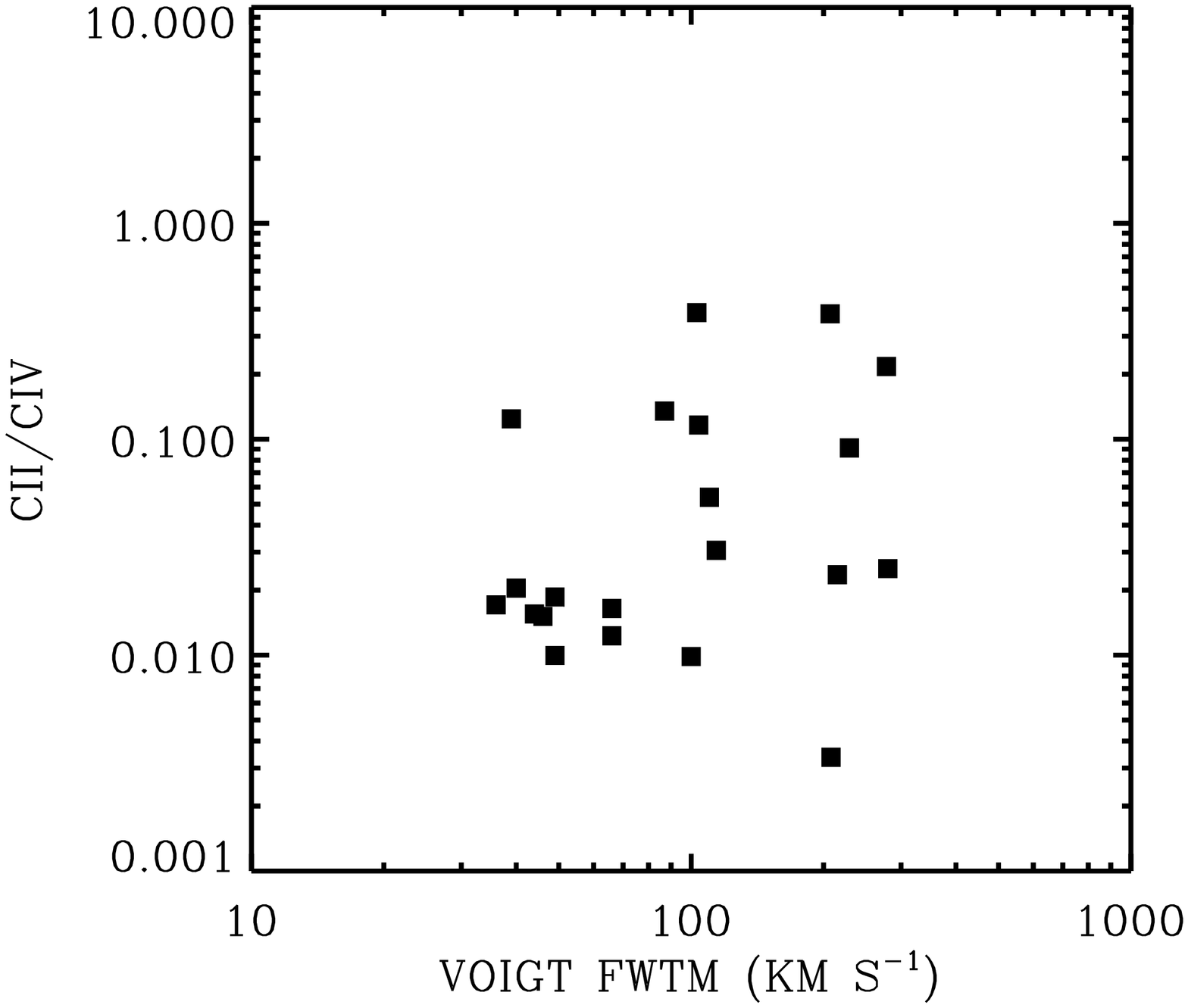}
  \includegraphics[width=2.8in,angle=90,scale=1.0]{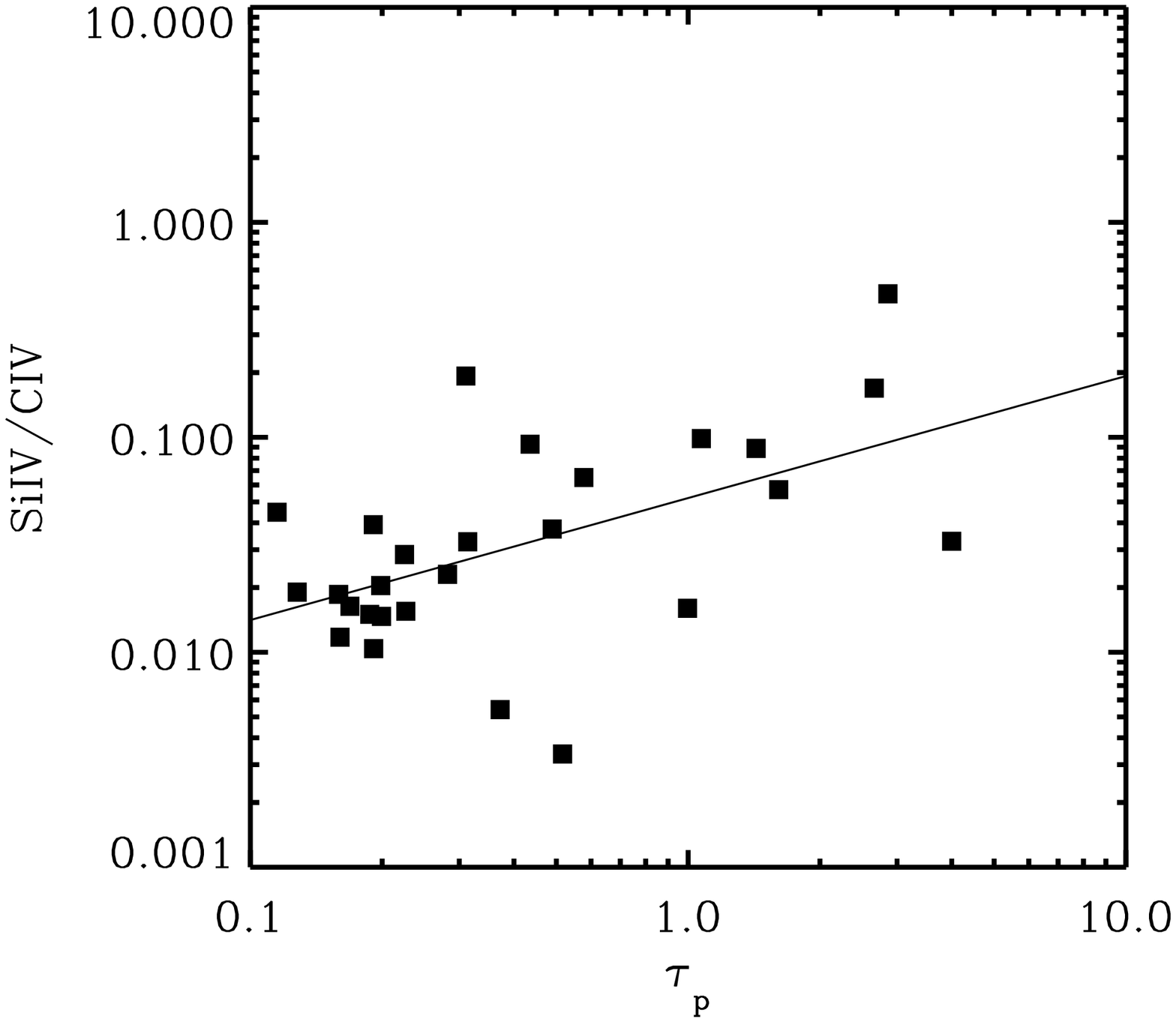}
  \includegraphics[width=2.8in,angle=90,scale=1.0]{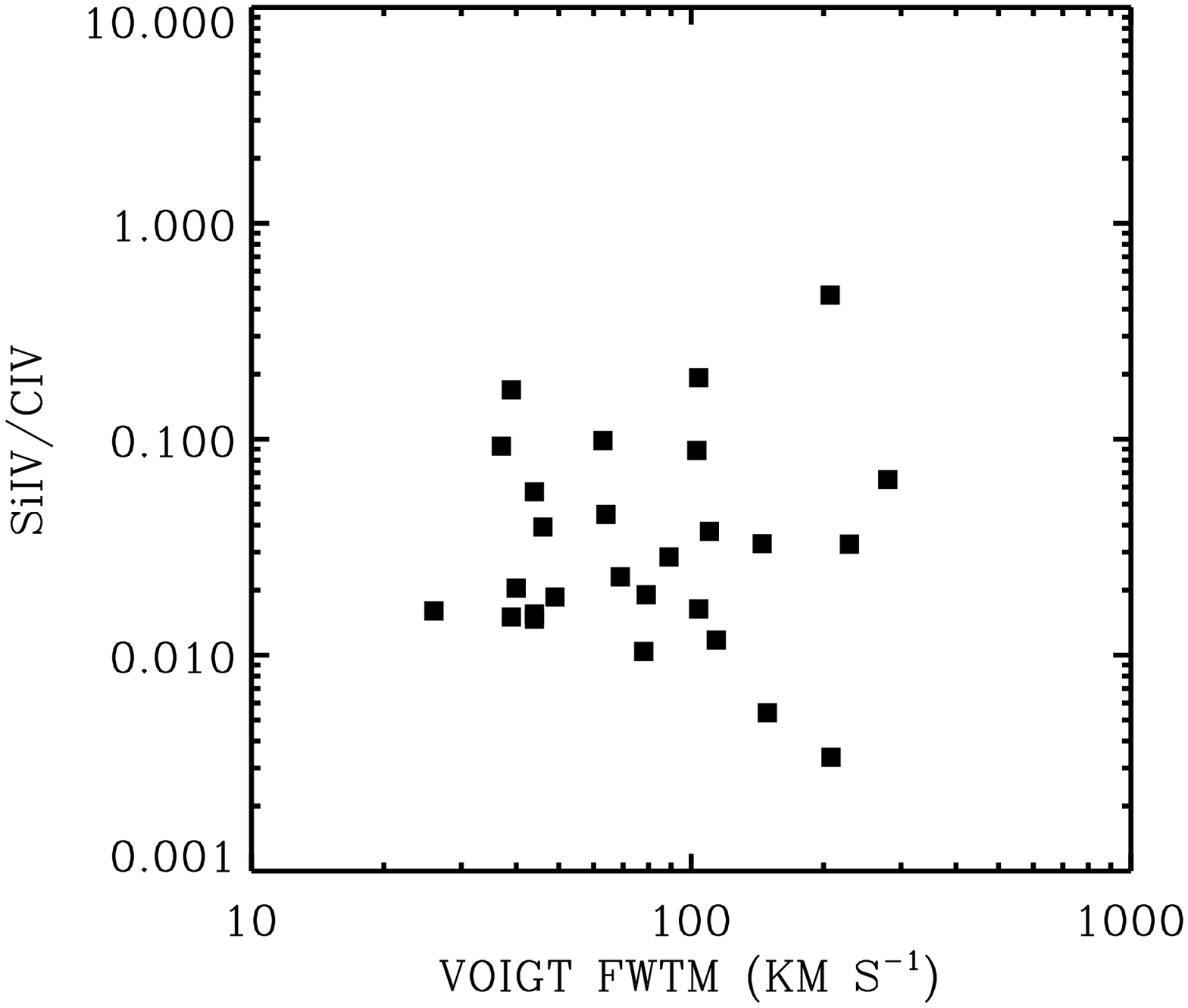}
  \caption{Ion ratios versus peak \ion{C}{4} 1548 optical depth and
  FWTM velocity width.  {\it Solid lines}: simple power law fits to
  the dependence of the ion ratios on peak optical depth.
  \ion{C}{2}/\ion{C}{4} shows a near-linear dependence on $\tau_p$,
  whereas \ion{Si}{4}/\ion{C}{4} shows a shallower rise.
\label{fig:c2c4_ta}
}
\end{figure}

 In Figure~\ref{fig:c2c4_ta} we show the dependence of
\ion{C}{2}/\ion{C}{4} and \ion{Si}{4}/\ion{C}{4} as a function of
$\tau_{p}$ and of $dv({\rm FWTM})$. \ion{C}{2}/\ion{C}{4} shows an
almost linear relation with $\tau_{p}$\ (though with some spread)
reflecting the fact that higher column density aborbers have lower
ionization parameters (e.g.\ Shaye et al.\ 2003).
\ion{Si}{4}/\ion{C}{4} shows a shallower but smooth rise with
$\tau_{p}$.  The near linearity of \ion{C}{2}/\ion{C}{4} is a fairly
natural consequence of a model in which the absorbers arise in the
general IGM since \ion{C}{2}/\ion{C}{4} is roughly dependent on the
inverse of the ionization parameter or, for a fixed ionzation, the
absorber density, whereas the \ion{C}{4} column density (or
$\tau_{p}$) is a rough measure of the abosorber column density which
in turn roughly follows the overdensity of the system. However, this
uniformity of behaviour would be much harder to understand in a model
in which the bulk of the absorbers arose in a galactic outflow and
where individual absorber properties might be expected to depend on
the strength of the outflow, the hydrodynamic convection and
fragmentation, and the properties of the galaxy itself, as well as any
galaxy contribution to the ionization.

 The \ion{C}{2}/\ion{C}{4} ratio plotted versus velocity width does
show a suggestion of a transition at $dv=100~{\rm km\ s^{-1}}$:
systems with lower velocity spreads are almost invariably
high-ionization systems with low \ion{C}{2}/\ion{C}{4} whereas the
wide systems show much more of a range in \ion{C}{2}/\ion{C}{4}
values. Consistent with the velocity analysis, this might suggest that
wide velocity systems could comprise a mixture of galaxy outflow and
IGM systems. However, the number of systems is too small for the
abruptness of the tranisition to be a robust conclusion and the data
could be consistent with a smooth evolution of \ion{C}{2}/\ion{C}{4}
with $dv$, where this would be a simple corollary of the density
dependence discussed in the previous paragraph.

\section{Conclusions}
\label{secconc}

The primary result of this paper is that weak \ion{C}{4} absorption
systems ($\tau_p < 0.4$\ or $N({\rm C~IV}) < 2 \times 10^{13}~{\rm
cm}^{-2}$) almost all lie in narrow complexes with a median FWTM of
$50~{\rm km\ s}^{-1}$.  It appears that these systems cannot be easily
understood in terms of high-velocity galactic-wind outflows from
galaxies that are active at that time.  Because there are so few wide
systems in the low-$\tau_p$\ sample (fewer than 10\%), the covering
factor would have to be much less than 1 and the bubble size and
outflow velocity unreasonably large when compared with observations of
Lyman break galaxies.  These systems also appear to be
ionized by an AGN-like spectrum.  As many as half of the stronger
\ion{C}{4} systems are wide enough that they could be produced in
local outflows and in some cases could be ionized primarily by the
parent galaxy, so that as many as half of higher column density
systems could be associated with contemporary outflows from galaxies.

\acknowledgments 

This research was
supported by the National Science Foundation under grant AST 00-98480.

\newpage


\clearpage
\begin{deluxetable}{ccccccccc}
\small
\tablewidth{450pt}
\tablenum{1}
\tablecaption{The C~IV Sample \label{tbl:1}}
\tablehead{
\colhead{Number} & \colhead{Quasar} & \colhead{$z_{peak}$}  & 
\colhead{$\tau_{peak}$}  & \colhead{$dv$} & \colhead{$z_{cen}$}  
& \colhead{$N_{\rm CIV}$} & \colhead{$N_{\rm SiIV}$} & \colhead{$N_{\rm CII}$}  
}
\startdata
 1   & HS1700+6416   & 2.31484   & 4.00   & 183   & 2.31536   & 1.9e+15   & 6.1e+13   & 0.0e+00   \nl
 2   & Q0636+680     & 2.47528   & 3.25   & 244   & 2.47416   & 1.7e+14   & 0.0e+00   & 0.0e+00   \nl
 3   & Q0636+680     & 2.90459   & 2.86   & 212   & 2.90374   & 3.0e+14   & 1.4e+14   & 1.2e+14   \nl
 4   & HS1700+6416   & 2.16793   & 2.66   &  42   & 2.16795   & 1.0e+14   & 1.7e+13   & 1.3e+13   \nl
 5   & HS0741+4741   & 2.58354   & 1.88   &  99   & 2.58334   & 1.0e+14   & 0.0e+00   & 0.0e+00   \nl
 6   & HS0741+4741   & 2.69237   & 1.61   & 245   & 2.69235   & 6.5e+13   & 3.7e+12   & 0.0e+00   \nl
 7   & Q1422+2309    & 3.53865   & 1.56   & 272   & 3.53863   & 1.8e+14   & 3.7e+13   & 4.0e+13   \nl
 8   & HS0741+4741   & 2.73297   & 1.43   & 121   & 2.73323   & 4.8e+13   & 4.3e+12   & 1.9e+13   \nl
 9   & Q0636+680     & 3.01724   & 1.27   & 317   & 3.01749   & 4.1e+13   & 1.1e+12   & 7.0e+11   \nl
10   & HS0741+4741   & 2.64916   & 1.07   &  63   & 2.64921   & 5.3e+13   & 5.2e+12   & 0.0e+00   \nl
11   & Q1422+2309    & 2.74886   & 1.00   & 113   & 2.74891   & 4.1e+13   & 0.0e+00   & 0.0e+00   \nl
12   & HS1946+7658   & 2.64432   & 1.00   &  23   & 2.64437   & 2.3e+13   & 3.7e+11   & 0.0e+00   \nl
13   & HS0741+4741   & 3.01778   & 0.85   & 117   & 3.01763   & 8.4e+13   & 2.4e+13   & 1.6e+15   \nl
14   & Q0636+680     & 2.31119   & 0.63   & 157   & 2.31098   & 7.2e+13   & 0.0e+00   & 0.0e+00   \nl
15   & Q1422+2309    & 3.38159   & 0.61   & 204   & 3.38277   & 4.8e+13   & 3.1e+12   & 1.1e+12   \nl
16   & HS1700+6416   & 2.43869   & 0.58   & 280   & 2.43863   & 5.2e+13   & 3.4e+12   & 1.3e+12   \nl
17   & Q0636+680     & 2.89183   & 0.52   & 215   & 2.89163   & 3.0e+13   &-1.7e+11   & 4.5e+09   \nl
18   & Q1422+2309    & 3.08354   & 0.50   & 245   & 3.08354   & 4.1e+11   & 0.0e+00   & 0.0e+00   \nl
19   & HS1700+6416   & 2.57802   & 0.49   & 110   & 2.57857   & 3.6e+13   & 1.3e+12   & 1.9e+12   \nl
20   & HS1700+6416   & 2.37988   & 0.44   &  31   & 2.37984   & 1.2e+13   & 1.1e+12   & 0.0e+00   \nl
21   & Q1422+2309    & 2.96195   & 0.39   & 155   & 2.96228   & 3.4e+13   & 0.0e+00   & 0.0e+00   \nl
22   & HS0741+4741   & 2.67287   & 0.37   & 107   & 2.67283   & 1.8e+13   &-5.7e+11   & 0.0e+00   \nl
23   & HS0741+4741   & 3.06665   & 0.35   &  59   & 3.06656   & 2.0e+13   & 3.1e+11   & 3.3e+11   \nl
24   & Q1422+2309    & 2.69835   & 0.33   & 221   & 2.69795   & 3.1e+13   & 0.0e+00   & 0.0e+00   \nl
25   & HS0741+4741   & 2.90527   & 0.31   & 232   & 2.90455   & 2.4e+13   & 7.8e+11   & 2.2e+12   \nl
26   & HS1946+7658   & 2.89309   & 0.31   & 101   & 2.89271   & 2.5e+13   & 4.9e+12   & 2.9e+12   \nl
27   & Q1422+2309    & 3.51459   & 0.30   &  44   & 3.51473   & 1.0e+13   & 1.4e+12   &-3.3e+11   \nl
28   & Q1422+2309    & 2.89538   & 0.29   &  69   & 2.89509   & 1.3e+13   & 0.0e+00   & 0.0e+00   \nl
29   & HS0741+4741   & 2.62110   & 0.28   & 112   & 2.62123   & 1.7e+13   & 3.8e+11   & 0.0e+00   \nl
30   & Q1422+2309    & 3.13440   & 0.25   & 357   & 3.13413   & 1.6e+13   & 1.1e+12   & 0.0e+00   \nl
31   & HS1946+7658   & 2.39519   & 0.25   &  25   & 2.39528   & 5.7e+12   & 0.0e+00   & 0.0e+00   \nl
32   & HS0741+4741   & 3.05360   & 0.23   &  19   & 3.05366   & 8.1e+12   & 1.3e+13   &-2.8e+11   \nl
33   & Q1422+2309    & 2.97619   & 0.23   & 366   & 2.97621   & 8.1e+12   & 0.0e+00   & 0.0e+00   \nl
34   & Q0636+680     & 2.86874   & 0.23   &  92   & 2.86883   & 6.4e+12   & 2.9e+10   &-3.9e+11   \nl
35   & HS1700+6416   & 2.02108   & 0.22   &  35   & 2.02108   & 8.3e+12   & 2.4e+11   & 0.0e+00   \nl
36   & Q0636+680     & 2.32485   & 0.22   &  52   & 2.32492   & 9.3e+12   & 0.0e+00   & 0.0e+00   \nl
37   & Q1422+2309    & 2.68275   & 0.22   &  49   & 2.68257   & 7.1e+12   & 0.0e+00   & 0.0e+00   \nl
38   & HS1700+6416   & 2.28955   & 0.20   &  49   & 2.28952   & 6.8e+12   & 6.2e+10   & 0.0e+00   \nl
39   & HS1946+7658   & 2.77725   & 0.20   &  35   & 2.77731   & 4.9e+12   &-1.3e+11   &-4.7e+11   \nl
40   & Q0636+680     & 2.68207   & 0.19   & 204   & 2.68228   & 9.6e+12   &-1.5e+11   & 0.0e+00   \nl
41   & HS1700+6416   & 2.56826   & 0.19   &  45   & 2.56817   & 6.6e+12   & 2.6e+11   &-5.5e+10   \nl
42   & HS1700+6416   & 2.19884   & 0.19   &  54   & 2.19893   & 6.7e+12   & 5.2e+10   & 0.0e+00   \nl
43   & Q1422+2309    & 2.72015   & 0.19   &  33   & 2.72009   & 6.1e+12   & 0.0e+00   & 0.0e+00   \nl
44   & HS1700+6416   & 2.12775   & 0.17   &  30   & 2.12778   & 6.1e+12   &-3.3e+11   & 0.0e+00   \nl
45   & HS0741+4741   & 2.96511   & 0.16   &  94   & 2.96530   & 8.5e+12   &-4.4e+11   & 2.6e+11   \nl
46   & HS1946+7658   & 2.91663   & 0.16   &  49   & 2.91661   & 5.4e+12   &-1.6e+11   & 6.7e+10   \nl
47   & Q1422+2309    & 3.41150   & 0.15   &  78   & 3.41121   & 9.3e+12   & 1.5e+12   & 1.3e+12   \nl
48   & Q1422+2309    & 2.94748   & 0.13   & 330   & 2.94754   & 3.0e+12   & 0.0e+00   & 0.0e+00   \nl
49   & HS1946+7658   & 2.65414   & 0.13   &  66   & 2.65424   & 7.8e+12   & 1.5e+11   & 0.0e+00   \nl
50   & HS0741+4741   & 3.03513   & 0.13   &  93   & 3.03469   & 1.0e+13   & 9.7e+11   & 5.9e+10   \nl
51   & HS0741+4741   & 2.71463   & 0.12   &  46   & 2.71446   & 6.3e+12   & 2.8e+11   & 0.0e+00   \nl
52   & Q0636+680     & 2.62142   & 0.11   &  58   & 2.62115   & 6.7e+12   & 0.0e+00   & 0.0e+00   \nl
53   & Q1422+2309    & 2.99918   & 0.11   &  28   & 2.99922   & 4.7e+12   & 0.0e+00   & 0.0e+00   \nl

\enddata

\end{deluxetable}

\newpage
\clearpage

\end{document}